\documentclass[aps,twocolumn,groupedaddress,showpacs,showkeys,nofootinbib,amsfonts,eqsecnum,floatfix]{revtex4}

\usepackage{graphicx}
\usepackage{epsfig}

\newcommand{\PP}{{\rm P}}
\newcommand{\NP}{{\rm NP}}
\newcommand{\tra}{\Delta}

\newcommand{\gb}{{\text{gb}}}
\newcommand{\betaa}{{\gamma}}
\newcommand{\betag}{{\beta}}
\newcommand{\alphac}{{\alpha}}
\newcommand{\R}{{\tilde{\alpha}}}
\newcommand{\comments}[1]{} 

\def\slashchar#1{\setbox0=\hbox{$#1$}
   \dimen0=\wd0 \setbox1=\hbox{/} \dimen1=\wd1
   \ifdim\dimen0>\dimen1 \rlap{\hbox to \dimen0{\hfil/\hfil}} #1
   \else  \rlap{\hbox to \dimen1{\hfil$#1$\hfil}} / \fi}

\begin{document}

\title{Trace anomaly, thermal power corrections \\ and dimension two
  condensates in the deconfined phase}

\author{E. Meg\'{\i}as}
\email{emegias@quark.phy.bnl.gov}

\affiliation{
Nuclear Theory Group, Physics Department,
Brookhaven National Laboratory,
Upton, New York 11973 USA
}

\author{E. \surname{Ruiz Arriola}}
\email{earriola@ugr.es}

\author{L.L. Salcedo}
\email{salcedo@ugr.es}

\affiliation{
Departamento de F\'{\i}sica At\'omica, Molecular y Nuclear,
Universidad de Granada,
E-18071 Granada, Spain
}

\date{\today}

\begin{abstract}
  The trace anomaly of gluodynamics on the lattice shows clear
  fingerprints of a dimension two condensate above the phase
  transition. The condensate manifests itself through even
  powers of the  inverse temperature while the total perturbative contribution
  corresponds to a mild temperature dependence and turns out to be
  compatible with zero within errors.  We try several resummation
  methods based on a renormalization group improvement. The trace
  anomaly data are analyzed and compared with other determinations of
  the dimension two condensate based on the Polyakov loop and the
  heavy $q\bar{q}$ free energy, yielding roughly similar numerical
  values. The role of glueballs near the transition is also discussed.
\end{abstract}



\maketitle


\section{INTRODUCTION}
\label{sec:intro}

The physics of the Quark-Gluon Plasma has turned out to be more
difficult than initially expected (for a recent review see
e.g. \cite{Shuryak:2008eq} and references therein). While it has been
noted~\cite{Lichtenegger:2008mh} that the failure of the perturbative
description could have been anticipated on general grounds, it has not
been obvious what is the form and furthermore the dynamical origin of
such a strongly non-perturbative behaviour. On the other hand it is by
now clear from lattice calculations at finite temperature (for a
review on recent developments see e.g.~\cite{DeTar:2008qi} and
references therein) that while in QCD there is a cross-over, in pure
Yang-Mills theory or gluodynamics for $N_c \ge 3 $ there is a first
order phase transition. Actually, in the simpler case of gluodynamics
the deconfinement phase transition is monitored by the expectation
value of the Polyakov loop which acts as an order parameter of the
associated breaking of the center ${\mathbb Z}(N_c)$ symmetry.

The finding of power corrections above the phase
transition~\cite{Megias:2005ve} and the possible explanation in terms
of a dimension-2 condensate of the dimensionally reduced theory of
such a special object suggests pursuing the same idea in other
thermodynamic quantities. Power corrections are ubiquitous at high
energies and zero temperature through condensates being the remnant of
non-perturbative effects.  From this viewpoint there is no reason why
they should not be present at high temperatures as genuinely
non-perturbative effects. In the present paper we detect and analyze
power corrections on the trace anomaly density at finite temperature
as obtained in lattice calculations, and which can only be effectively
explained by a dimension two condensate. Unlike the Polyakov loop, the
trace anomaly corresponds to a physical and measurable quantity not
only on the lattice but also experimentally in the QCD case. In this
way we extend our study of the non-perturbative effects above the
phase transition started for the Polyakov loop and the $q\bar{q}$ free
energy ~\cite{Megias:2005ve,Megias:2006ke,Megias:2007pq}. A brief
account of our results have been presented in
Refs.~\cite{Megias:2008dv,Megias:2008rm}.

The paper is organized as follows.  In Section~\ref{sec:pce3p} we provide
mounting evidence on the dominance of power corrections for the trace
anomaly using available lattice data for
gluodynamics~\cite{Boyd:1996bx}. The thermodynamics of the phase
transition is analyzed in Section~\ref{sec:4} following an interesting
insight by Pisarski~\cite{Pisarski:2006yk}. The physics below the
phase transition in the confined region should be describable in terms
of colour singlet glueball states, an issue that we address within a
large $N_c$ perspective where the interactions among glueballs are
much suppressed. In passing we note a Hagedorn looking glueball
exponentially growing spectrum with practically no impact on the
confined region. We analyze in 
Section
\ref{sec:RG-improvement}
our results from
the point of view of analytical schemes based on perturbation theory.
Actually, a common deficiency of the poorly converging perturbation
theory is the occurrence of unpleasant infrared divergences and
manifest spurious scale dependence. Perturbative resummations such as
Hard Thermal Loop remove the infrared singularities due to Debye
screening but provide too weak a signal as compared to the lattice
data. Our analysis suggests that those methods might in fact describe
the deconfined phase {\it after} the power corrections have been
subtracted off.  To reinforce this conclusion 
a renormalization
group improvement of the perturbative free energy is undertaken in this Section.
We find that besides implementing
renormalization scale independence, this resummation mimics the Hard
Thermal Loop calculations within estimated uncertainties, and as a
consequence similar conclusion follows: power corrections dominate the
deconfined phase while perturbatively based physics plays a minor role
wherever it may be considered applicable.
The dimension two condensate has naturally
explained the presence of power corrections for the Polyakov loop and
singlet $q\overline{q}$ free
energy~\cite{Megias:2005ve,Megias:2007pq}.  The idea is tried out in
Section~\ref{sec:dim2ctraceanomaly} for the trace anomaly where we
indeed show that there is a satisfactory numerical agreement for the
dimension two condensate with that of the Polyakov loop and the
$q\overline{q}$ free energy. Finally in Section~\ref{sec:conclusions}
we give our main conclusions.
In Appendix ~\ref{sec:2} we rederive
the trace anomaly at finite temperature for reference.
In Appendix ~ \ref{subsec:4.C} the possible Hagedorn pattern of the
glueball spectrum is analyzed. In Appendix ~\ref{app:3} we study an improvement of the perturbative expansion of the free energy at finite temperature based on renormalization group invariance.

\section{Thermal power corrections in the trace anomaly}
\label{sec:pce3p}

For zero or infinite quark masses (gluodynamics) the classical scale
invariance of  QCD is broken by quantum corrections due to the
necessary regularization yielding a trace
anomaly~\cite{Collins:1976yq}. In the context of QCD sum rules
phenomenology the issue has been discussed in
Ref.~\cite{Novikov:1981xj} (for a review see
e.g. \cite{Shifman:1988zk}). At finite temperature the trace of the
energy momentum tensor is related not only to the trace anomaly but
also to the difference $ \epsilon - 3 P $ with $\epsilon$ the energy
density and $P$ the pressure~\cite{Landsman:1986uw} (see also
Ref.~\cite{Leutwyler:1992ic,Leutwyler:1992mv}) providing a measure of
the interaction as well as the anomalous breaking of scale
invariance. An early lattice calculation of both energy and pressure
was first undertaken one decade ago~\cite{Boyd:1995zg,Boyd:1996bx}
(see also more recent
calculations~\cite{Okamoto:1999hi,Gavai:2005da,Umeda:2008bd,Umeda:2008kz})
exhibiting the expected deconfinement phase transition although far
still from a free gluonic plasma. Finite temperature Ward identities
and low energy theorems based on the trace anomaly have been deduced
in Refs.~\cite{Ellis:1998kj,Shushpanov:1998ce} in the continuum and
extended to Euclidean lattices~\cite{Meyer:2007fc}. Renormalization
issues and the identity of thermodynamic and hydrostatic pressure was
discussed in Ref.~\cite{Drummond:1999si} within the real time
formulation. Phenomenological implications in the context of resonance
gas model are analyzed in \cite{Agasian:2001bj}. A good knowledge of
the trace anomaly is crucial to understand the deconfinement process,
where the non perturbative (NP) nature of low energy QCD seems to play
a prominent role~\cite{Lichtenegger:2008mh}.

The derivation of the well known relation
\begin{eqnarray}
\epsilon - 3 P = \frac{\betag(g)}{ 2 g } \langle
(G_{\mu\nu}^a)^2 \rangle  \,,
\label{eq:anomaly}
\end{eqnarray}
between the trace anomaly and the beta function of gluodynamics is
reviewed in Appendix \ref{sec:2}.

\begin{figure}[tbp]
\begin{center}
\epsfig{figure=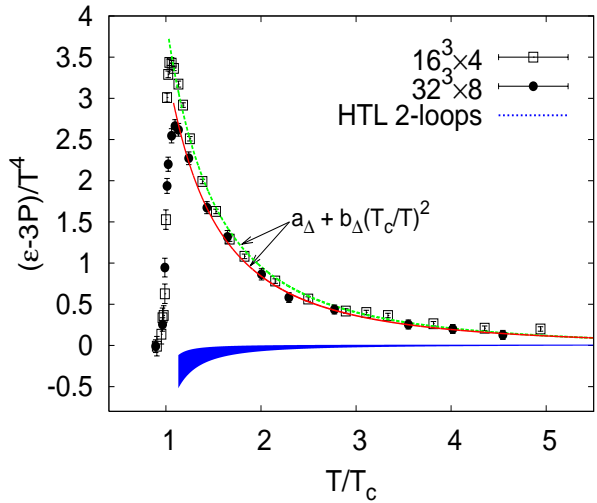,height=6.5cm,width=8.5cm}
\end{center}
\caption{The trace anomaly density $ (\epsilon -3 P)/T^4 $ as a
  function of $T$ (in units of $T_c$ ). Lattice data are
  from~\cite{Boyd:1996bx} for $N_\sigma^3 \times N_\tau = 16^3 \times
  4$ and $32^3 \times 8$. The fits using Eq.~(\ref{eq:fit}) are
  plotted. 2-loop result for the trace anomaly density in HTL
  perturbation theory, from \cite{Andersen:2002ey}, is shown as a
  shaded band that correspond to varying $\mu$ by a factor of two
  around $2\pi T$.}
\label{fig:e3p_1}
\end{figure}

The trace anomaly density has been computed at finite temperature on
the lattice for pure gluodynamics~\cite{Boyd:1996bx}. The standard
plot of 
\begin{eqnarray}
\Delta \equiv \frac{\epsilon -3 P}{T^4} = T \frac{\partial}{\partial
T} \left( \frac{P}{T^4} \right)  \label{eq:deltatermo}
\end{eqnarray}
as a function of $T$ is shown in Fig.~\ref{fig:e3p_1}. Below the
critical temperature the trace anomaly $\Delta $ is very small though
not exactly zero. $\Delta$ increases suddenly near and above $T_c$ by
latent heat of deconfinement, and raises a maximum at $T \approx 1.1
\, T_c$. Then it has a gradual decrease reaching zero in the high
temperature limit. The high value of $\Delta$ for $T_c \le T \le 3\,
T_c$ corresponds to a strongly interacting Quark-Gluon Plasma picture.

\begin{figure}[tbp]
\begin{center}
\epsfig{figure=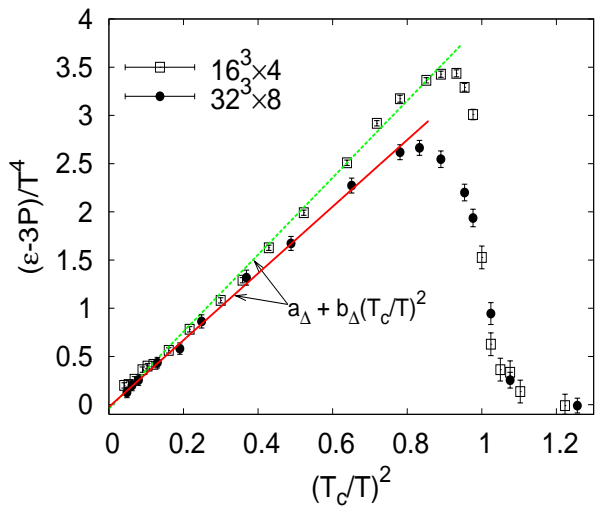,height=6.5cm,width=8.5cm}
\end{center}
\caption{The trace anomaly density $ (\epsilon -3 P)/T^4 $ as a
  function of $1/T^2$ (in units of $T_c$ ).
Same data as in Fig.~\ref{fig:e3p_1}.
}
\label{fig:e3p_2}
\end{figure}

In previous works~\cite{Megias:2005ve,Megias:2007pq} we have detected
the presence of inverse power corrections on other thermal
observables. Guided by our previous experience we represent in
Fig.~\ref{fig:e3p_2} the dimensionless interaction measure $(\epsilon
-3 P)/T^4 $ as a function of $1/T^2$ (in units of $T_c$). This plot
exposes an unmistakable straight line behaviour corresponding to a
power correction of the form
\begin{eqnarray} 
\frac{\epsilon - 3 P}{T^4} = a_\tra + b_\tra \left(
\frac{T_c}{T} \right)^2
\label{eq:fit} 
\end{eqnarray} 
in the deconfined phase region slightly above the critical
temperature. A direct fit of the lattice data $(N_\sigma^3 \times
N_\tau = 16^3 \times 4) $ for $1.13 < T/T_c < 5.0 $ yields
\begin{equation}
a_\tra = -0.041(17), \quad b_\tra= 3.99(5), \quad \chi^2/
{\rm dof} = 5.0 \nonumber \\
\label{eq:ab_exp1}
\end{equation} 
with correlation parameter $r(a,b)=-0.78$. Likewise, the fit for
$(N_\sigma^3 \times N_\tau = 32^3 \times 8)$ and $1.13 < T/T_c < 4.5 $
yields
\begin{equation}
a_\tra = -0.02(4), \quad b_\tra= 3.46(13), \quad \chi^2/
{\rm dof} = 0.35 \nonumber \\
\label{eq:ab_exp}
\end{equation} 
with $r(a,b)=-0.73$. The fit is more accurate in the
second set of lattice data. This set corresponds to a more precise
determination of the trace anomaly on the lattice. In order to
estimate the result in the continuum limit, we can assume that the
difference between the two lattice results is entirely due to finite
cutoff effects. Assuming further that the corresponding leading effect
goes as $1/N^2_\tau$, yields the estimate $a_\tra=-0.02(9)$
and $b_\tra= 3.28(27)$ in the continuum limit.

A fit of the data completely excludes the existence of the odd power
corrections $T_c/T$ and $\left(T_c/T\right)^3$ in $\Delta$. We have
attempted to determine the coefficient of a possible quartic
correction, appending formula~(\ref{eq:fit}) with a term $c_\tra
\left(T_c/T\right)^4$. A fit of lattice data for $N_\tau =8$ results
in $a_\tra=-0.05 (7) $, $ b_\tra=3.7 (7)$ and 
$c_\tra=-0.3(9)$ with $\chi^2/ {\rm dof} = 0.31 $ and correlations
$r(a,b)=-0.85$ , $r(b,c)=-0.97$ and $r(a,c)=0.75$. So, more accurate
data are desirable in order to identify contributions from condensates
of dimension 4. 

In any case it is clear the existence of power corrections which are
beyond the scope of the radiative corrections accounted for in
perturbation theory. It should be mentioned that similar power
corrections have been identified in the trace anomaly in gluodynamics
in $2+1$ dimensions ~\cite{Bialas:2008rk}.

Integrating the trace anomaly, as modeled in (\ref{eq:fit}), from
$T=T_c$ and using the continuity of the pressure across the phase
transition yields the following form for the pressure
\begin{eqnarray}
\frac{P(T)}{T^4}= \frac{b_\tra}{2} \left(1-\frac{T_c^2}{ T^2}\right)
 + a_\tra \log \left(\frac{T}{T_c} \right)  
+  \frac{P(T_c)}{T_c^4} . \nonumber \\
\end{eqnarray}
Clearly this expression cannot be used at very high temperatures
unless $a_\tra=0$, a result suggested by the fit. On the opposite extreme,
integrating from very high energies, where radiative corrections should
dominate, yields
\begin{eqnarray}
\frac{P(T)}{T^4}= \frac{b_0}2  -  \int_{T}^\infty \frac{dT'}{T'} \Delta_{\rm
  pert} (T')
\end{eqnarray}
where 
\begin{equation}
  b_0 = (N_c^2-1) 2\pi^2/45 \approx 3.51
  \,.
\end{equation}
Matching at some temperature $T=T_p$, gives $b_\tra$ from
$P_c=P(T_c)$, $b_0$ and $\Delta_{\rm pert}(T)$ (assuming $a_\tra=0$).
Clearly, since $P_c>0$, and $\Delta_{\rm pert} (T)>0 $, we expect
$b_\tra < b_0$. (Note that the power correction should be negligible
at the matching temperature.)  The result from the fit almost
saturates the inequality since the continuum limit extrapolation
yields $b_\tra=3.28(27)$ which suggests that {\it both} $P_c$ and
$\Delta_{\rm pert} (T) $ must be very small. The analysis of
Section~\ref{sec:4} confirms these issues.

\section{Thermodynamics of the phase transition}
\label{sec:4}

The analysis of the previous section leaves little doubt on the
existence of power corrections in gluodynamics down {\it almost} to
the critical temperature. Moreover, at the maximum temperature
measured on the lattice, $T=5 \,T_c$, the power contribution still
dominates the result, and both terms become comparable only at higher
temperatures, $T \sim 10\, T_c$. One sensible question to ask is what is
the temperature where perturbation theory applies. We address this
question in more detail in Section~
\ref{sec:RG-improvement}.

If for the moment perturbative corrections are disregarded (they are
small anyhow), it is worth reviewing why the previous results are
completely against the idea of a gluon plasma.  Actually, the standard
textbook argument requires the bag pressure to equilibrate the free
gluon gas at the critical temperature
\begin{eqnarray}
P(T)= P_{\rm gluons}(T)-B , \qquad T > T_c, 
\end{eqnarray}
yielding the trace anomaly 
\begin{eqnarray}
\Delta= \frac{4 B}{T^4}, \qquad T > T_c .
\end{eqnarray}
As already shown, this behaviour is excluded by lattice data.
 
In this section we elaborate on a suggestion by
Pisarski~\cite{Pisarski:2006yk}, 
\begin{equation}
P(T)=\frac{b_0}{2}(T^4-T_c^2T^2),
\end{equation}
which achieves a fair
description of the data assuming infinitely heavy glueball masses, 
negligible radiative perturbative corrections, and a temperature
dependent bag energy, called {\it fuzzy bag},
\begin{eqnarray}
B_{\rm fuzzy} = \frac{b_0}2 T_c^2 T^2
.
\end{eqnarray}
As shown in 
Section~\ref{sec:dim2ctraceanomaly},
dimension-2 condensates provide
a natural explanation of this fuzzy bag picture {\it including} the
correct power behaviour and coefficient, while simultaneously account
for power corrections in the Polyakov loop and the
$q\bar{q}$-potential at a quantitative level. A study of the
thermodynamics of this fuzzy constant is also made in
Ref.~\cite{Andreev:2007zv} within the holographic QCD approach based
on the AdS/CFT correspondence.

\subsection{Infinitely massive glueballs}

At very high temperatures, where radiative corrections become
negligible, the free gluonic gas relation $P (T) \sim (N_c^2-1) \pi^2
T^4 /45$ holds. On the other hand, in the confined phase, i.e., below
the critical temperature $T_c$, the spectrum is believed to be
saturated by glueballs~(for reviews see e.g.
Ref.~\cite{Teper:1998kw,Mathieu:2008me}). Since the pressure must be a
continuous function across the phase transition, one finds 
\begin{equation}
P(T_c)=P_{\rm
  glueballs} (T_c)
.
\end{equation}
However, the lightest glueball has $J^{PC}=0^{++}$ and $M_{0^{++}}=
1.73(1) {\rm GeV}$. This is much heavier than the critical temperature
$T_c \approx 270\,{\rm MeV}$ (a mysteriously disparate
scale~\cite{Ishii:2003en}). Thus $\Delta $ is very small (though not
exactly zero) in the confined phase. Taking the infinite glueball mass
limit produces $P_{\rm glueballs}=0$.  These conditions are met by the
interpolating functions of the form
\begin{eqnarray}
P= \frac{(N_c^2-1) \pi^2}{45} T^4 \left[ 1- \left( \frac{T_c}{T}
  \right)^n \right] \, , \qquad T \ge T_c \, , 
\end{eqnarray} 
with arbitrary positive $n$. This parameterization yields
\begin{eqnarray} 
\Delta = \frac{n(N_c^2-1) \pi^2}{45} \left( \frac{T_c}{T} \right)^n \,,
\qquad T \ge T_c \, , 
\label{eq:e3pnfit}
\end{eqnarray} 
which for $n=2$ corresponds to take $a_\tra=0$ and $b_\tra=3.51$ in
Eq.~(\ref{eq:fit}), in excellent agreement with the fit to the lattice
data of the previous section. This thermodynamic consistency does not
explain however {\it why} there is a power correction with $n=2$. A
fit of Eq.~(\ref{eq:e3pnfit}) to the data of Fig.~\ref{fig:e3p_2} for
the same range as in Eq.~(\ref{eq:ab_exp}), $1.13 T_c \le T \le 4.54 T_c
$, yields $n=1.97(5)$ with a slightly larger $\chi^2/{\rm dof}=0.62$.

\subsection{Finite mass glueball spectrum}

Given the goodness of the description we may try to improve by
actually computing the pressure of an ensemble of finite mass
glueballs in the confined phase using the 12 glueball spectrum
currently determined on the
lattice~\cite{Morningstar:1999rf,Chen:2005mg}. Quite generally such a
calculation would require the use of the quantum field theoretical
version of the virial expansion ~\cite{Dashen:1969ep} and thus to take
into account corrections from binary, ternary and higher order
collisions. For instance, for a binary collision the threshold for the
first possible lowest glueball-glueball resonance is located at twice
the glueball mass $2 M_{0^{++}} \sim 3 {\rm GeV}$ which lies above the
5th glueball and is expected to produce an insignificant contribution
for $T < T_c \approx 270\, {\rm MeV}$. Fortunately, the first four
low-lying glueballs computed on the lattice are kinematically stable
bound states. Actually, within a large $N_c$ framework one has a gas
of stable and non-interacting glueballs since $M_\gb \sim N_c^0$ and
$V_{\gb\text{-}\gb} \sim 1/N_c^2$. So it seems safe to treat the
glueballs as a gas of free bosons. This yields
\begin{eqnarray}
P_{\rm glueballs}(T) 
&=& 
\frac{1}{3} \sum_i g_i \int \frac{d^3
  k}{(2\pi)^3} \frac{\vec k \cdot \nabla_k E_k }{e^{E_k/T}-1}
,
\label{eq:p-glueballs} 
\end{eqnarray}
where the index $i$ runs on the glueball species: $M_i$ denotes the
glueball mass and $g_i=2J_i+1$ the corresponding angular momentum
degeneracy. A convenient low temperature expansion gives
\begin{eqnarray}
P_{\rm glueballs}(T) 
=
\sum_i \sum_{n=1}^ \infty \frac{g_iM_i^2 T^2}{2 n^2 \pi^2 }
K_2  \! \left( \frac{M_i n}{T} \right)
\label{eq:p-glueballs1} 
\end{eqnarray}
where the sum over $n$ corresponds to the
thermal loops and $K_2$ is a modified Bessel function of the second kind.

Using known spectrum of the lightest 12 glueball
species~\cite{Morningstar:1999rf,Chen:2005mg}  one can evaluate
$P_c = P_{\rm glueball} (T_c)$. This requires fixing the critical
temperature scale $T_c$ where the trace anomaly is
evaluated~\cite{Boyd:1996bx} on the one hand and the Sommer scale
$r_0$ defined as $r_0^2 V_{q\bar q}'(r_0) =
1.65$~\cite{Sommer:1993ce}, where the glueball spectrum is
obtained~\cite{Morningstar:1999rf,Chen:2005mg}, on the other. This is
necessary since the numerical value of $P_c$ is quite sensitive to the
glueball mass to critical temperature ratio, $M_i/T_c$. 

Boyd {\it et al.}~\cite{Boyd:1996bx} find $T_c/\sqrt{\sigma}=0.629(3)$
while Luscher {\it et al.}~\cite{Luscher:1993gh} obtain $r_0
\sqrt{\sigma}=1.22(8)$. This is consistent with the elaboration of
Teper~\cite{Teper:1998kw} where $r_0 \sqrt{\sigma}=1.195(10)$, $T_c
/\sqrt{\sigma}=0.640(15)$ and $M_{0^{++}} /\sqrt{\sigma}=3.52(11)$ are
quoted as well as the fixed point action of Niedemeyer {\it et
  al.}~\cite{Niedermayer:2000yx} simulation where the values $r_0
\sqrt{\sigma}=1.197(11)$, $T_c /\sqrt{\sigma}=0.624(7)$, $r_0 T_c
=0.750(5)$ and scalar $r_0 M_{0^{++}}=4.12(21)$ and tensor $r_0 M_{2^{++}}
=5.96(24)$ low-lying glueball masses. Along similar lines
Necco~\cite{Necco:2003vh}, gets $r_0 T_c =0.7498(50)$. These latter
values agree within errors with the 12 glueball spectrum
simulation~\cite{Morningstar:1999rf,Chen:2005mg} where $r_0 M_{0^{++}}
=4.21(11)$ and $r_0 M_{2^{++}} =5.85(2)$ vs $r_0 M_{0^{++}}=4.16(11)$
and $r_0 M_{2^{++}} =5.85(5)$ are obtained respectively. Using for
definiteness $r_0 T_c=0.75$ we get the result
\begin{eqnarray}
P_c = P_{\rm glueball} (T_c) = 0.01(1)
\end{eqnarray}
where the error has been estimated from the uncertainty in the
glueball masses. This value of the pressure is consistent with the
fit to the data, although the glueball spectrum contribution is not so
strongly constrained by the trace anomaly data. About $80\%$ of the
total contribution to $a_\tra $ is given by the lightest scalar
$0^{++}$ and tensor $2^{++}$ glueball states, confirming the marginal
role of the excited glueball spectrum as well as possible two-glueball
resonances.

The trace anomaly due to the glueballs reads 
\begin{eqnarray}
\Delta(T) = 
\sum_i \sum_{n=1}^ \infty \frac{g_i}{2 n \pi^2 } \frac{M_i^3}{T^3} 
K_1\! \left( \frac{M_i n}T \right) 
.
\label{eq:em3p-glueballs} 
\end{eqnarray}
At the lowest lattice temperature $T=0.89\, T_c$ one has $\Delta <
0.08$. This sets an upper bound since all contributions in
Eq.~(\ref{eq:em3p-glueballs}) are positive. Taking the known glueball
spectrum one gets $\Delta_\gb (0.89 T_c)= 0.03(6)$. However, the steep
raise below $T_c$ is not reproduced by the glueball spectrum,
basically due to the heavy values of the masses. In this context it
has been suggested~\cite{Datta:2002je} that above the phase transition
glueballs lower their masses providing an enhancement already below
the phase transition.
The behaviour and effect of glueballs in the deconfined phase has also
been considered in \cite{Brau:2009mp}.
Note also that the glueball gas formula neglects
glueball interactions which are nominally $1/N_c^2$.  Assuming that
the multiplicity in the glueball spectrum is ${\cal O}(N_c^0)$ this
yields a trace anomaly $\Delta ={\cal O}(N_c^0)$ for $T < T_c$, while
we clearly have $\Delta ={\cal O}(N_c^2)$ for $T > T_c$.

Further considerations on the glueball spectrum and its possible
Hagedorn structure are presented in Appendix \ref{subsec:4.C}.
In all, our analysis above explains why the pressure of finite mass
glueballs is compatible with zero within errors both from the trace
anomaly lattice calculation as well as from the direct glueball
spectrum estimate, and hence supports the infinite mass glueball
assumption underlying the model of Ref.~\cite{Pisarski:2006yk}.



\section{Renormalization group improvement of the perturbative free energy}
\label{sec:RG-improvement}

The weak coupling
expansion for the free energy of the quark-gluon plasma has been
calculated through order
$\alpha_s^{5/2}$~\cite{Arnold:1994ps,Zhai:1995ac,Braaten:1995ju}, and
the result yields poor convergence at the lattice QCD available
temperatures $T < 5 T_c$, and even at much higher temperatures. The
problem is more involved taking into account that at order $\alpha_s^3
\log \alpha_s$ starts the contribution of non perturbative effects
which is hard to compute~\cite{Kajantie:2002wa}. There have also been
numerous attempts to resum perturbation theory in order to get a
better convergence of the result (see e.g. \cite{Andersen:2004fp} and
references therein).  One of the most developed techniques is the Hard
Thermal Loop (HTL) perturbation theory, first proposed
in~\cite{Braaten:1989mz,Frenkel:1989br}. This is an efficient
reorganization of the perturbative series which avoids the unpleasant
infrared singularities. We show in Fig.~\ref{fig:e3p_1} the two-loop
result for the trace anomaly density from HTL~\cite{Andersen:2002ey}
(see also \cite{Andersen:1999sf} for a one loop computation). The HTL
result becomes a very smooth function in the regime $T > 1.13\, T_c$
and is unable to reproduce the lattice data. One of the most
conspicuous facts is that the pressure from HTL remains nearly
constant with a value of about $95\%$ of that of an ideal gas of
gluons, even for temperatures $\sim 10^3\, T_c$. This means that the
approach to the ideal gas is extremely slow, which ultimately implies
that the trace anomaly is very small and smooth. The same feature is
shared by the weak coupling expansion and other resummation
techniques, and it seems to be confirmed by preliminary results of
lattice computations at extremely high
temperatures~\cite{Endrodi:2007tq}.

Many perturbative studies based on the computation of the
three-dimensional effective theory of QCD on the lattice lead once
again to the same smooth behaviour for the trace anomaly even near
$T_c$~\cite{Kajantie:2000iz,Hietanen:2008tv}.

\begin{figure}[tbp]
\begin{center}
\epsfig{figure=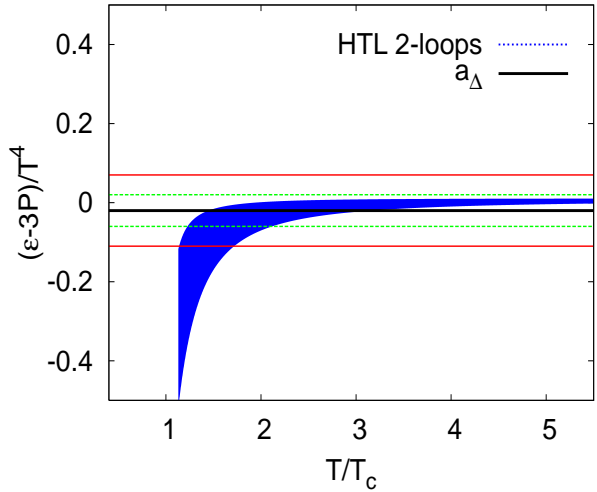,height=6.5cm,width=8.5cm}
\end{center}
\caption{The trace anomaly density $ (\epsilon -3 P)/T^4 $ as a
  function of $T$ (in units of $T_c$). The 2-loop HTL result from
  ref.~\cite{Andersen:2002ey} is shown as a shaded band that
  corresponds to varying the scale $\mu$ by a factor of two around
  $2\pi T$.  The horizontal lines display the fitted value of $a_\tra$
  in Eq.~(\ref{eq:fit}), from $N_\tau = 8$ lattice data, $a_\tra=
  -0.02\pm 0.04$ (error bar in green), and the continuum limit estimate, $a_\tra= -0.02\pm 0.09$ (error bar in red).
}
\label{fig:e3p_3}
\end{figure}

These problems can be understood within our framework by noting that
perturbation theory, as well as HTL and other resummation techniques,
contain only logarithms in the temperature, suggesting a mild
temperature dependence. Our discussion above shows that these
approaches would yield a powerless contribution, $\Delta_{\rm pert}$,
which should ultimately be identified with the {\it almost} constant
and vanishing $a_\tra$ of Eq.~(\ref{eq:fit}) rather than with the full
result from the lattice. Actually, the maximum lattice temperature $T
= 5\, T_c$ should be far from the pQCD estimate since the power
correction provides the bulk of the full result at this temperature.
For $N_c = 3$ the ${\cal O}(g^5)$ corrections to Eq.~(\ref{eq:e3pPT})
corresponds to multiply it by $(1-6 g(\mu)/\pi)$~\cite{Braaten:1995jr}
which becomes small for $g(\mu) \ll \pi/6$ or $\mu \gg 10^{11}
\Lambda_{\rm QCD}$. This delayed onset of perturbative QCD is not a
new phenomenon and takes place in the study of exclusive processes at
high energies where there appear collinear divergences (see e.g.
Ref.~\cite{RuizArriola:2008sq} and references therein). In order to
support this statement, we zoom in Fig.~\ref{fig:e3p_3} the two-loop
result for the trace anomaly density in HTL and compare it with our
value $a_\tra = -0.02(9)$ from a fit of lattice data using
Eq.~(\ref{eq:fit}), estimated in the continuum limit. The HTL result
is inside the error band for temperatures $T > 1.4(3)\, T_c$. From a
fit of $N_\tau = 8$ lattice data we got $a_\tra = -0.02(4)$, and the
HTL result is inside the error band for $T > 1.7(4) \, T_c$. It is
appropriate to mention at this point that the estimated error of the
2-loop HTL start to be important below $1.5 \, T_c$ as we can see in
Fig.~\ref{fig:e3p_3}. Moreover below $1.5 \, T_c$ the 1-loop and
2-loop computations in HTL start to be in
disagreement~\cite{Andersen:2002ey}.

There are in the literature other computations of the equation of
state of gluodynamics from lattice simulations that are in qualitative
agreement with Boyd et al. (see
Ref.~\cite{Okamoto:1999hi,Gavai:2005da,Umeda:2008bd,Umeda:2008kz} and
references therein). Power corrections have also been obtained in the
equation of state of full QCD~\cite{Cheng:2007jq}. Nonetheless we will
focus in this paper only on pure gluodynamics for simplicity.
 
The conclusion of the above discussion is that the behaviour of
Eq.~(\ref{eq:fit}) clearly contradicts perturbation theory which
contains no powers but only logarithms in the temperature.  Power
corrections are understood as the high energy trace of
non-perturbative low energy effects. This kind of corrections have already
been detected and analyzed in the Polyakov loop and in the heavy
quark-antiquark free energy in terms of dimension two gluon
condensates~\cite{Megias:2005ve,Megias:2006ke}. The rather successful
description of the lattice data for the trace anomaly suggests a
similar approach for the equation of state, that will be analyzed in
Sec.~\ref{sec:dim2ctraceanomaly}.

Our previous numerical analysis of lattice data suggests that the
perturbative series of the free energy of a hot gluon plasma is rather
small, not only in the high temperature limit, but also near the
deconfined phase. The perturbative series has a problem of convergence
for temperatures below $ 10\, T_c$, which could mean a lack of
analyticity near the phase transition. 

We can improve the
convergence by considering a resummation of the perturbative series
based on renormalization group invariance. 
The details of the computation are provided in Appendix~\ref{app:3}.
In practice this means to
reorganize the perturbative series and express it in terms of the
following manifestly Renormalization Group (RG) invariant
\begin{equation}
\R(T) = \frac{1}{\betaa_0} \frac{1}{\log\left(\frac{ 2 \pi
T}{\Lambda_{\rm QCD}}\right)} .
\label{eq:RGinv}
\end{equation}
The result for the pressure so obtained is a manifestly RG invariant series up to order ${\cal  O}(\R^3)$ and is given by Eq.~(\ref{eq:pRG}) with coefficients
Eq.~(\ref{eq:RGcoef}). The only unknown coefficient to this order is
$A_6$, which enters at ${\cal O}(\R^3)$. The trace anomaly involves
a derivative of the pressure with respect to temperature, see
Eq.~(\ref{eq:deltatermo}). Using (\ref{eq:RGinv}), this can be computed
as
\begin{equation}
\Delta_{\rm pert} = 
-\frac{(N_c^2-1)\pi^2}{45}\betaa_0\R^2\frac{d H_{\rm pert}}{d\R}
\,, \label{eq:DeltaPT}
\end{equation}
where we have defined $H_{\rm pert} = P_{\rm
  pert}/P_{\rm ideal}$, being $P_{\rm ideal} = ((N_c^2-1)\pi^2/45)
T^4$ the pressure of an ideal gas of massless gluons.
This result holds modulo ${\cal O}(\R^{9/2})$ when Eq.~(\ref{eq:pRG})
is considered. 

One can try to compare the RG-invariant $\Delta_{\rm pert}$, truncated
to ${\cal O}(\R^{7/2})$ (included) with the available lattice data. In
all cases we use the more accurate $N_\tau=8$ data, including points
in the range $1.13\le T/T_c\le 4.5$. To begin with we allow both $A_6$
and $\Lambda_{\rm QCD}$ to change as free parameters. In this form no
acceptable fit is achieved. The best fit gives $\chi^2/{\rm dof}=1.6$
with $A_6=0.33\pm 3.0$. It mimics the overall shape of the data but
fails to reproduce the highest temperature data, where perturbation
theory (PT) is expected to be more reliable.  In addition, this best
fit requires unphysical values of $\Lambda_{\rm QCD}\sim 338\,{\rm
  MeV}$, which would not reproduce other gluodynamics data.  To avoid
this problem, in what follows we fix $\Lambda_{\rm QCD}$ to lie on the
range $0.877(88)\,T_c$, allowing for a $10\%$ uncertainty in this
parameter.  In this case, fitting $A_6$ in order to reproduce the
highest temperature data, one can verify that the $\chi^2/{\rm dof}$
rapidly deteriorates as new lower temperature data are being included
in the fit, while simultaneously the fitted value of $A_6$ rapidly
changes to larger negative values.

\begin{figure}[tbp]
\begin{center}
\epsfig{figure=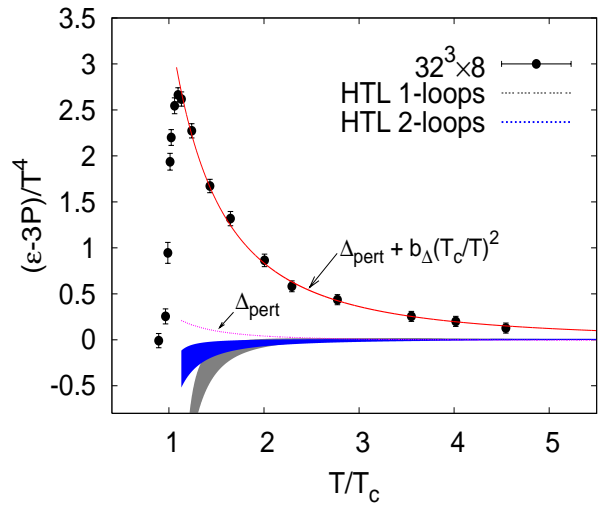,height=6.5cm,width=8.5cm
}
\end{center}
\caption{The trace anomaly density $(\epsilon - 3P)/T^4$ as a function
  of $T$ (in units of $T_c$). Lattice data are from \cite{Boyd:1996bx}
  for $N_\sigma^3\times N_\tau = 32^3 \times 8$. The fit using
  Eq.~(\ref{eq:PT+NP}) is plotted. We also plot the perturbative
  contribution from Eq.~(\ref{eq:DeltaPT}), and compare it with the
  1-loop \cite{Andersen:1999sf} and 2-loop \cite{Andersen:2002ey} HTL
  results (shaded bands).  }
\label{fig:PT+NP}
\end{figure}

\begin{figure}[tbp]
\begin{center}
\epsfig{figure=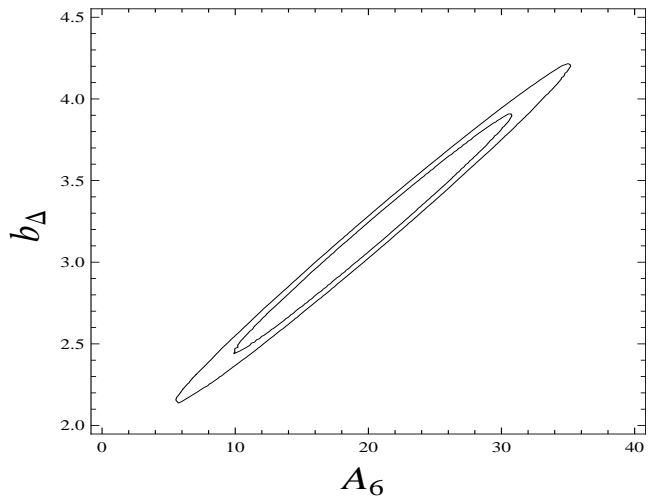,height=6.5cm,width=8.5cm
}
\end{center}
\caption{Correlation ellipses corresponding to $\Delta \chi^2 = 2.3, 4.7$
(${\rm dof}=6$) between the perturbative parameter $A_6$
and the non perturbative parameter $ b_\Delta$ (see main text)
characterizing the trace anomaly. The central values are obtained using
the lattice data in the interval $1.13 T_c \le T \le 4.54 T_c$.}
\label{fig:ellipse}
\end{figure}

Next we proceed to add the non perturbative (NP) term
$b_\tra(T_c/T)^2$ to the perturbative ones and try to reproduce the
data using $A_6$ and $b_\tra$ as parameters. In a more detailed
treatment one should expect some interference between these
contributions, from radiative corrections in the form of anomalous
dimension, etc, in the NP term.  Presently, we adopt the simplest
scenario of additive PT and NP contributions, i.e. we consider
\begin{equation}
\Delta = \Delta_{\rm pert} + b_{\Delta} \left(\frac{T_c}{T}\right)^2 \,. 
\label{eq:PT+NP}
\end{equation} 
This procedure yields a fairly good fit to the data, $\chi^2/{\rm
dof}=0.40$, with $b_\tra=3.18(74)$ and $A_6=20.0\pm 10.5$ (see
Fig.~\ref{fig:PT+NP}).  This fit is only slightly worse than that in
section \ref{sec:pce3p}, which used a constant perturbative
background $a_\tra$. The error bars in $b_\tra$ and $A_6$ are enhanced as they
include the uncertainty in $\Lambda_{\rm QCD}$. It is also noteworthy
that the two parameters are highly correlated (see
Fig.~\ref{fig:ellipse}), and in fact the combination $A_6-14.2\,
b_\Delta=-25.1\pm 1.3$  (which has zero correlation with $b_\tra$) has a much smaller error. Nevertheless, we
caution that these numbers are to be taken as indicative only. The missing
terms in the perturbative expansion make the perturbative contribution
to be little reliable at temperatures near the transition. For
instance, if the previous fit is repeated using data in the range
$1.43 \le T/T_c \le 4.5$, the central value of $b_\tra$ increases to
$3.47$ while $A_6$ changes correspondingly.

A remarkable fact is that the central value of $A_6$ is such the
$\Delta_{\rm pert}$ falls on top of the 2-loop HTL calculation.
Actually, the value of $A_6$ required to produce this match between
the two perturbative calculations at $T=4.5\,T_c$ is $A_6=19.0\pm 1.2$
(note that this latter number does not depend on lattice trace anomaly
data). Similar values, $A_6= 20.0\pm 0.8$, allow to reproduce the HTL
result by $\Delta_{\rm pert}$ in the range $2.75\le T/T_c \le 4.5$
with $\chi^2/{\rm dof}=0.67$ (where we include the uncertainty from
$\Lambda_{\rm QCD}$ and from the scale $\mu$ in HTL). While the
central values of $A_6$ obtained from a fit to lattice data, including
the NP term, and to 2-loop HTL are equal ($20.0$), the uncertainty
assigned from the second determination is much smaller.
(Nevertheless, the preferred value of $A_6$ as extracted from a match
to HTL evolves at very high temperatures to a value close to zero.)

We have also considered a fit to the lattice trace anomaly data using
the 2-loop HTL result as perturbative background, plus the non
perturbative term, in the range $1.24\le T/T_c \le 4.5$. This gives
$b=3.62(11)$ with $\chi^2/{\rm dof}= 0.72$. So it also provides a good
fit of the data, only slightly worse than that from the RG invariant
$\Delta_{\rm pert}$. 

\begin{figure}[tbp]
\begin{center}
\epsfig{figure=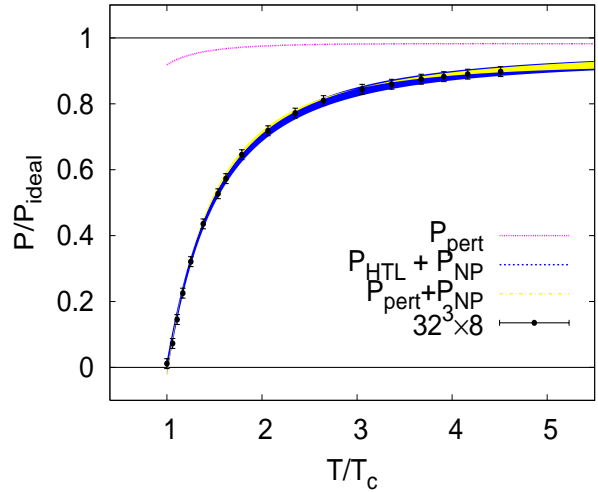,height=6.5cm,width=8.5cm
}
\end{center}
\caption{The pressure $P$ as a function of $T$ (in units of $T_c$). Lattice data are from \cite{Boyd:1996bx} for $N_\sigma^3 \times N_\tau = 32^3 \times 8$. We plot the fit of Eq.~(\ref{eq:pPT+NP}), and the analogous equation with $P_{\rm HTL}$ to 2-loops~\cite{Andersen:2002ey}. We retain the same values of $b_\Delta$ as in Fig.~\ref{fig:PT+NP}. The perturbative contribution with no extra contributions is also displayed. }
\label{fig:pressure}
\end{figure}

In all cases the perturbative contribution is subdominant and very
small above $2\,T_c$. Also, the pressure of gluodynamics is very well
reproduced for all temperatures above $T_c$, after inclusion of the non
perturbative term. Integrating the trace anomaly, as modeled in (\ref{eq:PT+NP}), yields the following form
\begin{equation}
\frac{P(T)}{T^4} = c_p +  \frac{P_{\rm pert}(T)}{T^4} - \frac{b_{\Delta}}{2}\left(\frac{T_c}{T}\right)^2 \,, \label{eq:pPT+NP}
\end{equation}
where $c_p$ is an additive constant related to the normalization of
the thermodynamic quantities on the lattice. Note that $P_{\rm
  pert}(T)$ is known from Eq~(\ref{eq:pRG}). A fit of the lattice data
for the pressure using Eq.~(\ref{eq:pPT+NP}) is plotted in
Fig.~\ref{fig:pressure}. We retain in the fit the same value of
$b_{\Delta}$ that we previously obtained from the trace anomaly, and
consider as free parameters $A_6$ and $c_p$. This formula yields a
good fit to the data in the regime $ 1 \le T/T_c \le 4.5$ with $A_6 =
20 \pm 10$, $c_p = -0.08 \pm 0.08$ and $\chi^2/{\rm dof} = 0.26$. As
expected, the value of $A_6$ agrees with our previous determination
from the trace anomaly. We also consider a fit of the lattice data
using the 2-loops HTL results for the pressure, $P_{\rm
  HTL}$~\cite{Andersen:2002ey}. The only free parameter in this case
is $c_p$, and we get $c_p = 0.006 \pm 0.020$, $\chi^2/{\rm dof} =
0.96$. We have increased the errors for the pressure of
Ref.~\cite{Boyd:1996bx} by a factor $2.5$, and it seems to be reasonable
taking into account the difference between lattices with different
temporal extent $N_\tau$.

We have also analyzed different lattice data other than those primarily
considered above~\cite{Boyd:1996bx}, namely Ref.~\cite{Okamoto:1999hi}
which almost reproduces~\cite{Boyd:1996bx} and
Ref.~\cite{Umeda:2008bd} which differ from those of
Ref.~\cite{Boyd:1996bx} in the higher temperature region, possibly due
to systematic errors not provided in Ref.~\cite{Umeda:2008bd}. An
estimate of the systematic errors based on a variation of lattice
sizes suggests that they could be of the order of $0.3$ in $\Delta$,
which explains the discrepancy of $0.15$ between both lattice
data. Thus, we have artificially increased the errors of
Ref.~\cite{Umeda:2008bd} by a factor of ten. For the resummed PT+NP
approach we get $A_6 = 31(7) $ and $b_\Delta = 3.9(0.5)$ with $
\chi^2/{\rm dof} =0.40 $ for the data of Ref.~\cite{Okamoto:1999hi}
while $A_6 = -9(18) $ and $ b_\Delta = 1.42(1.45) $ with $\chi^2/{\rm
dof}=0.12 $ when data from Ref.~\cite{Umeda:2008bd} are used. 
Note that the resulting pairs $(A_6,b_\tra)$ fall reasonably close to the 
line $A_6-14.2\, b_\tra=-25.1\pm 1.3$. 
The
HTL+NP yields $b_\Delta = 3.44(0.09) $ with $ \chi^2/{\rm dof}=2.6$
for Ref.~\cite{Okamoto:1999hi}, and $b_\Delta = 4.05(0.15) $ with $
\chi^2/{\rm dof}=0.44$ for Ref.~\cite{Umeda:2008bd}. With the provisos
spelled out previously, it is quite reasonable to conclude that other
lattice data sets confirm our results, with larger uncertainties.

To finish this section we note that the trace anomaly as obtained from
$\Delta_{\rm pert}$ or from HTL does not have a definite sign, but
this quantity becomes positive definite for all temperatures above
$T_c$ upon inclusion of the non perturbative term $b_\tra(T_c/T)^2$,
for values of $b_\tra$ fitting the data. Curiously, the fitted value
$A_6=20$ is roughly the threshold  value above which $\Delta$ would fail to be
positive definite (the region with negative values being located
around $T=50\,T_c$).

\section{Dimension two condensate and trace anomaly}
\label{sec:dim2ctraceanomaly}

We can make a first approximation to the computation of the trace
anomaly using the non perturbative model introduced in
~\cite{Megias:2005ve,Megias:2007pq}. There, the inverse temperature
power corrections observed in lattice data for the Polyakov loop
\cite{Kaczmarek:2002mc} and the $q\bar{q}$-potential
\cite{Kaczmarek:2005ui}, are accounted for by means of a non
perturbative contribution to the zeroth component gluon propagator of
the dimensionally reduced theory. Namely,
\begin{equation}
D_{00}(\mathbf{k}) = D_{00}^{\PP}(\mathbf{k}) + D_{00}^{\NP}(\mathbf{k}) \,,
\label{eq:D00}
\end{equation}
where  
\begin{eqnarray}
D_{00}^{\rm P}({\mathbf k}) = \frac{1}{{\mathbf k}^2 + m_D^2} 
,
\quad
D_{00}^{\NP}(\mathbf{k}) = \frac{m_G^2}{(\mathbf{k}^2 + m_D^2)^2} \,.
\label{eq:D00NP}
\end{eqnarray}
Here $m_D$ is the Debye mass, which scales like $T$, and $m_G$ is a
temperature independent mass parameter.  The non perturbative term in
the gluon propagator induces a corresponding contribution to the
dimension two gluon condensate:
\begin{equation}
\langle A_{0,a}^2\rangle^{\NP} = \frac{(N_c^2-1)T m_G^2}{8\pi m_D}
.
\label{eq:A0NP}  
\end{equation}

Being the trace anomaly in
gluodynamics given by Eq.~(\ref{eq:anomaly}), the problem can be
addressed through the computation of the vacuum expectation value of
the squared field strength tensor $\langle (G_{\mu\nu}^a)^2\rangle$.

In perturbation theory $\langle G^2_{\mu\nu,a} \rangle$ starts at
${\cal O}(g^2)$, cf. (\ref{eq:e3pPT}). This comes from integration of
{\it ``hard modes''}, i.e., modes with momentum scale $~2\pi T$
~\cite{Braaten:1996jr}.  The contribution of soft modes, described by
the dimensionally reduced Lagrangian, starts at ${\cal O}(g^3)$. These
perturbative contributions break scale invariance through radiative
corrections and so depend logarithmically on the temperature, after
extracting the canonical factor $T^4$. Power corrections require a
stronger breaking of scale invariance. We will assume that this does
not happen for the hard modes since condensates are low energy
phenomena. Therefore, we assume that the non perturbative scale
breaking enters through the parameter $m_G^2$ in (\ref{eq:D00NP})
corresponding to the temporal gluon propagator. Just by dimensional
counting, this parameter will produce the required power correction,
$\sim 1/T^2$ in the observable $\Delta$. The ultraviolet divergent
momentum integrals in three dimensions are dealt with by dimensional
regularization. On the other hand no non perturbative scale breaking
is assumed for spatial gluons. As these gluons fail to have a Debye
mass, a non perturbative term of the type $m_G^\prime{}^2/k^4$ in the
propagator of spatial gluons would yield a vanishing contribution
within dimensional regularization.  This approach yields
\begin{eqnarray} 
\langle G_{\mu\nu,a}^2 \rangle^{\NP} &=& 2 \langle
(\partial_i A_{0,a})^2 \rangle^{\NP}
\nonumber
\\
&=&
2 T(N_c^2-1)\int \frac{d^3k}{(2\pi)^3}k^2 D_{00}^\NP
\label{eq:8.1}
\\
&=&
- T(N_c^2-1)\frac{3}{4\pi}m_D m_G^2
.
\nonumber
\end{eqnarray} 
The negative sign is a consequence of the renormalization.\footnote{
  As advertised, an ${\cal O}(g^3)$ contribution, namely, proportional
  to $m_D^3$, is derived from the analogous computation using the {\em
    perturbative} component of the gluon propagator.} Making use of
(\ref{eq:A0NP}), this result can be expressed in a more conveniently
form in terms of the dimension two gluon condensate, and finally gives
\begin{equation}
(\epsilon - 3P)_\NP = 
-3 \frac{\betag(g)}{g} m_D^2 \langle A_{0,a}^2\rangle^\NP \,.
\label{eq:e3pfr}
\end{equation}
The behaviour is $\sim T^2$, in contrast to the perturbative behaviour
$\sim T^4$. It is also noteworthy that $(\epsilon - 3P)_\NP$
satisfies the correct large $N_c$ counting, being ${\cal O}(N_c^2)$.

The coefficient $b_\tra$ in the parameterization (\ref{eq:fit}) is to
be identified with the estimate of Eq.~(\ref{eq:e3pfr})\footnote{It
  may look unnatural that the rather complicated r.h.s. of this
  formula yields $b_\tra$, when Pisarski's argument, section
  \ref{sec:4}, assigned to this quantity an almost geometrical value,
  $(N_c^2-1)2\pi^2/45$. However, this is not so. In the present model,
  the geometrical value of $b_\tra$ together with the r.h.s. provide
  rather an estimate of $T_c$, identified as the temperature where the
  pressure becomes negligible.}
\begin{eqnarray}
b_\tra T_c^2 = -3\frac{ \betag (g) }{g} \frac{m_D^2}{T^2} \langle
A_{0,a}^2 \rangle^{\rm NP} \,.
\label{eq:bTc2}
\end{eqnarray} 
Therefore, since $\betag(g) < 0 $, we have a positive $b_\tra$. If we
consider the perturbative value of the beta function $\betag(g) \sim
g^3 + {\cal O}(g^5) $, the r.h.s. of Eq.~(\ref{eq:bTc2}) shows a
factor $g^2$ in addition to the dimension two gluon condensate $g^2
\langle A_{0,a}^2\rangle^{\rm NP}$. So the fit of the trace anomaly
data is sensitive to the value of $g$. Fortunately, this quantity has
only a smooth dependence on $T$. For the Polyakov loop the sensitivity
in $g$ is only through the perturbative terms, which are much smaller
than the NP ones. When we consider the perturbative value $g_P$ to
2-loop, $g_P\sim 1.41$, we get from the fit of the trace anomaly
\begin{eqnarray}
  g^2 \langle  A_{0,a}^2 \rangle^\NP 
  =
  ( 0.72(3) \; {\rm GeV} )^2
.
 \label{eq:A0NPvalue}
\end{eqnarray} 
Here we have used the continuum limit estimate $b_\tra=3.28(27)$ of
section \ref{sec:pce3p}, based on a constant perturbative background.
For the present purposes we consider this estimate sufficiently
accurate and less subject to details related to the precise
determination of the perturbative background.  The precise number for
$b_\tra$ varies by using any of the determinations of $b_\tra$
discussed in section \ref{sec:RG-improvement} but they are still
compatible with each other.

The value quoted in Eq.~(\ref{eq:A0NPvalue}) is a factor 1.5 smaller
than the value obtained from a fit of the Polyakov loop $(0.84(6) \;
{\rm GeV})^2$~\cite{Megias:2005ve} and heavy $q\overline{q}$ singlet
free energy $(0.90(5) \; {\rm GeV})^2$~\cite{Megias:2007pq}. This
disagreement could be explained in part on the basis of certain
ambiguity of $g$ in the non perturbative regime.  We show in
Ref.~\cite{Megias:2007pq} that a better fit of the Polyakov loop and
heavy quark free energy lattice data in the regime $T_c < T < 4\, T_c$
is obtained for a value of $g$ slightly smaller than $g_P$, i.e. $g =
1.26 - 1.41$. Taking this,  we get 
\begin{eqnarray}
g^2 \langle A_{0,a}^2 \rangle^{\rm NP} &=& (0.77(6) \; {\rm GeV})^2
\nonumber \\
&=&
(2.84 \pm 0.21 \, T_c)^2
\,,
\end{eqnarray}
in better agreement with determinations of the condensate from
previous observables, see Table~\ref{tab:cond}. This value follows
from the continuum limit estimate of $b_{\Delta}$, and its discrepancy
with the corresponding value from a fit of $N_\tau = 8$ lattice data
is $5 \%$, which is much smaller than the error. The same happens with
the lattice data of the Polyakov loop~\cite{Megias:2005ve}.

\begin{table}[htb]
\begin{center}
\begin{tabular}{|c|c|}
\hline
{\bf Observable} &  $\hspace{0.3cm} { g^2 \langle A_{0,a}^2 \rangle^{\rm NP} }$ \hspace{0.3cm}  \\
\hline
\hspace{0.3cm} Polyakov loop~\cite{Megias:2005ve} \hspace{0.3cm} &  \hspace{0.3cm} $(3.11 \pm 0.22 \, T_c)^2   $ \hspace{0.3cm} \\
\hspace{0.3cm} Heavy $\overline{q}q$ free energy~\cite{Megias:2007pq} \hspace{0.3cm}  &  \hspace{0.3cm} $(3.33 \pm 0.19 \, T_c)^2$  \hspace{0.3cm} \\ 
\hspace{0.3cm} Trace Anomaly  \hspace{0.3cm} &   \hspace{0.3cm} $(2.84 \pm 0.21 \, T_c)^2$ \hspace{0.3cm} \\ 
\hline
\end{tabular}
\end{center}
\caption{Values of the dimension two gluon condensate from a fit of
several observables in the deconfined phase of gluodynamics: Polyakov
loop, singlet free energy of heavy quark-antiquark and trace
anomaly. Values are in units of $T_c$. We show the fit for lattice
data with $N_\tau = 8$ for the heavy $\overline{q}q$ free energy, and the continuum limit estimate for the others. Error in last line takes into account an
indeterminate value of the coupling constant $g = 1.26 - 1.41$, being
the highest value the perturbative $g_P$ up to 2-loop at $T = 2\,
T_c$. The critical temperature in gluodynamics is taken as $T_c = 270 \pm 2
\,{\rm MeV}$~\cite{Beinlich:1997ia}.}
\label{tab:cond}
\end{table}



In a recent
paper  the electric-magnetic asymmetry of the condensate 
defined as
\begin{equation}
\Delta_{A^2}(T) =  
g^2 \langle A_{0,a}^2 \rangle - \frac{1}{3} g^2 \langle A_{i,a}^2\rangle,
\label{eq:9.1} 
\end{equation} 
is computed in SU(2) Yang-Mills theory~\cite{Chernodub:2008kf}.
Obviously this quantity vanishes at zero temperature by Euclidean
symmetry. The dimension two condensate has dimension of mass squared
and so it would vanish in a perturbative calculation at zero
temperature. At finite temperature instead $\langle A_{0,a}^2
\rangle^{\rm P}$ scales as $T^2$ modulo slowly varying radiative
corrections. On the other hand our model assumes that the non
perturbative part of the condensate $\langle A_{0,a}^2\rangle^\NP$ is
temperature independent (modulo radiative corrections) at least in the
regime slightly above the phase transition and beyond 
This is exactly the behaviour that is obtained
in~\cite{Chernodub:2008kf} for the asymmetry in this regime. If we
naively scale our determination for $g^2\langle A_{0,a}^2\rangle^\NP$
with the number of gluons $3/8$, we get in SU(2) the estimate
$g^2\langle A_{0,a}^2\rangle^\NP = (1.9(2) T_c)^2$.  This value can be
compared with the temperature independent contribution to
(\ref{eq:9.1}), $\Delta_{A^2}^{(0)} = (0.894(14)\, T_c)^2$
of~\cite{Chernodub:2008kf}, leaving some room for a magnetic
contribution in our model. In any case, the very existence of the
dimension two condensate seems to be confirmed by the work of
ref.~\cite{Chernodub:2008kf}.

\section{Conclusions} 
\label{sec:conclusions}

For quite a long time it has been believed that the highest
temperature measured in lattice calculations $T_{\rm max} \sim 5 \,
T_c$ could be matched smoothly to perturbative calculations. Assigning
the standard $\overline{\rm MS}$ conversion factor to a momentum scale
$\mu \sim 2 \pi T$ and taking $T_c \sim 0.6\sqrt{\sigma}$ yields
$\mu_{\rm max} \sim 1.5 \,{\rm GeV}$; a high scale. Actually, there
have been many perturbative studies that unsuccessfully tried to
reproduce the lattice data for the trace anomaly.  As we have shown in
this work, the trace anomaly density shows unmistakable traces of
power corrections in the inverse temperature in the region slightly
above the phase transition which may be accommodated by dimension two
gluon condensate. This finding parallels a similar analysis for the
Polyakov loop and the values for the dimension two condensate is in
satisfactory agreement with the value found here. The $20\%$ discrepancy may
be due to many different sources such as anomalous dimension effects.
It is important to emphasize that the non-triviality of such a
numerical consistency can only be addressed in a calculation where all
quantities can be estimated simultaneously. In addition, it would be
worrisome if the dimension two condensate associated with the power
correction would differ by an order of magnitude. The fact that this
is by far not the case suggests searching for other features where the
condensate might generate further genuinely non-perturbative thermal
power corrections. This of course applies not only to gluodynamics but
also to the case of full QCD with active dynamical quarks and in the
presence of a chemical potential.  A possible guideline for future
studies would be a simple minded extension of the power corrections
ideas discussed in the present work. Actually, most recent QCD
calculations already confirm the ubiquitous $1/T^2$ behaviour of the
trace anomaly density \cite{Cheng:2007jq}.

On a more methodological level, let us remind that at high energies
and zero temperature momentum power corrections are widely accepted
and encode via condensates non-perturbative effects. For this
situation the full machinery of QCD sum rules has been extensively
developed and applied with recurrent success. While there is no
reason why condensates should not be present at high temperatures as
genuinely non-perturbative effects even after deconfinement,
temperature power corrections above the phase transition are much less
common and a more complete technology than that used in the present
work would be most useful. In addition to the standard finite
temperature complications one must also face the well known ambiguity
on clear separation between perturbative effects from existing
non-perturbative features. There is no doubt that this intricate
theoretical problem should necessarily be addressed at some stage in
the near future.

In spite of the overall numerical consistency found for the dimension
two condensate, the physics underlying the generation of such a
condensate, and ultimately the very appearance of thermal power
corrections remains uncertain.  Actually, the elusiveness of a
theoretical determination of the dimension two condensate from first
principles reflects our current ignorance and remains a challenging
bottleneck to the understanding of strongly interacting quark gluon
plasma in non Abelian gauge theories.

\begin{acknowledgments}
E.~Meg\'{\i}as is supported from the joint sponsorship by the
Fulbright Program of the U.S. Department of State and Spanish Ministry
of Education and Science.  Work supported by Spanish DGI and FEDER
funds with grants no. FIS2008-01143/FIS, Junta de Andaluc\'{\i}a grant
FQM-225-05, EU Integrated Infrastructure Initiative Hadron Physics
Project contract RII3-CT-2004-506078, and U.S. Department of Energy
contract DE-AC02-98CH10886. We thank D. Kharzeev and A. Dumitru for
useful comments, and P. Petreczky for a careful reading of the
manuscript. We also thank S. Ejiri for providing us with the lattice
data of Ref.~\cite{Umeda:2008bd}.
\end{acknowledgments}

\appendix

\section{Trace anomaly at finite temperature}
\label{sec:2}

In this appendix we review the derivation of the trace anomaly at finite
temperature~\cite{Leutwyler:1992ic,Ellis:1998kj,Shushpanov:1998ce,Drummond:1999si,Agasian:2001bj}. In the Euclidean path
integral representation, the partition function for gluodynamics reads
\begin{eqnarray}
Z=\int {\cal D} \bar A_{\mu ,a} \exp \left[- \frac1{4 g^2} \int d^4 x
(\bar G_{\mu\nu}^a)^2 \right] 
\end{eqnarray}
where the field strength tensor is given by $\bar G_{\mu \nu} =
\partial_\mu \bar A_\nu - \partial_\nu \bar A_\mu - i [\bar A_\mu,
\bar A_\nu] $ and $(\bar{G}_{\mu \nu}^a)^2 > 0 $ in Euclidean space. Gauge
fixing terms will be considered in the gluonic measure ${\cal D} \bar
A_{\mu ,a}$, with periodic boundary conditions. The canonical gluon
fields are given by $A_\mu = \bar A_\mu /g$. This provides the relation
\begin{eqnarray}
\frac{\partial \log Z}{\partial g } = \frac{1}{2g^3}
\left\langle \int d^4 x
(\bar{G}_{\mu\nu}^a)^2 \right\rangle 
= \frac{1}{2g}\frac{V}{T}\langle  (G_{\mu\nu}^a)^2
\rangle
\label{eq:2.1a}
\end{eqnarray}
where in the last equality translational invariance has been used.

From the standard thermodynamic relations, the free energy, the
pressure and energy density are given by
\begin{eqnarray}
  F &=& -PV = - T \log Z \\  
  \epsilon &=& \frac{E}{V} =  \frac{T^2}{V} 
\frac{\partial  \log Z }{\partial T}
\end{eqnarray}  
as well as the relation
\begin{eqnarray}
\epsilon- 3 P  = T^5 \frac{\partial}{\partial T}
\left( \frac{P}{T^4} \right) \,. 
\label{eq:em3p}
\end{eqnarray}

The dimensionless quantity $P/T^4$ is a function of $T/\Lambda_{\rm
  QCD}$ (or rather $\Lambda_{\rm gluodynamics}$ in our case) and in
principle it can be expressed to all orders in terms of the coupling
constant and the scale:
\begin{eqnarray}
\frac{P}{T^4}= f(g(\mu),\log(\mu/2\pi T)).
\end{eqnarray}
The right-hand side is actually independent of the scale $\mu$. This allows to
compensate a variation in $T$ with a variation in $\mu$ and then in
$g$ (i.e., in $\Lambda_{\rm QCD}$)
\begin{eqnarray}
\frac{\partial}{\partial \log T}\left(\frac{P}{T^4}\right)
=
\frac{\partial g}{\partial\log\mu}\frac{\partial}{\partial g}
\left(\frac{P}{T^4} \right)
\end{eqnarray}

Using now
\begin{eqnarray}
\frac{P}{T^4}= \frac{\log Z}{VT^3}
\end{eqnarray}
with (\ref{eq:2.1a}) and (\ref{eq:em3p}) yields  finally the trace
anomaly,
\begin{eqnarray}
\epsilon - 3 P = \frac{\betag(g)}{ 2 g } \langle
(G_{\mu\nu}^a)^2 \rangle  \,,
\end{eqnarray}
where we have introduced the beta function
\begin{eqnarray}
\betag (g) = \mu \frac{d g}{d \mu } \,.
\end{eqnarray}

In this work we focus on the evaluation of the trace anomaly
density above the phase transition, which in practice means momenta
above $\mu \sim 1 \, {\rm GeV}$ if the customary $\overline{\rm MS}$
conversion factor $\mu= 2 \pi T $ is adopted. In perturbation theory
the beta function reads
\begin{eqnarray}
\betag (g) = - \betag_0 g^3  +  {\cal O}(g^5)
\end{eqnarray} 
where $\betag_0 = 11 N_c/(48 \pi^2)$.

The result for the trace anomaly density to two loops
is~\cite{Kapusta:1979fh}
\begin{eqnarray}
\frac{\epsilon- 3 P }{T^4 } &=& \frac{N_c(N_c^2 -1)}{72}
\betag_0 g^4(T) + {\cal O} ( g^5 )
\label{eq:e3pPT}
\end{eqnarray} 
where $1/g^2(\mu) = \betag_0 \log (\mu^2/\Lambda_{\rm QCD}^2)$ at
leading order. Remarkably this perturbative result implies a negative
expectation value of the squared field strength tensor; this quantity
being positive prior to renormalization.  Note the ambiguity in the
(truncated) perturbative result, since generally one has both the
temperature $T$ and the $\overline{\rm MS}$-renormalization scale
$\mu$, for which it is usually taken in the literature the reasonable
but $arbitrary$ choice $\mu \sim 2 \pi T$. This issue is analyzed in
section~\ref{sec:RG-improvement}. Higher order corrections including
up to $ g^6 \log g $ can be traced from~\cite{Kajantie:2000iz}.
Unfortunately severe infrared problems in the perturbative expansion
yield poor convergence at temperatures available to lattice QCD
calculations, $T < 5\, T_c$.

\section{The glueball Hagedorn spectrum}
\label{subsec:4.C}

\begin{figure}[tbp]
\begin{center}
\epsfig{figure=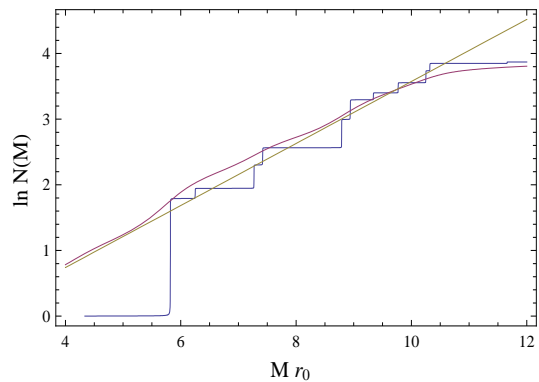,height=5cm,width=7cm}
\end{center}
\caption{Logarithmic plot of the cumulative number of glueball states
  obtained from the lattice as a function of the mass in units of the
  Sommer parameter $r_0$ compared to the exponential spectrum fit
  $N(M) = A e^{M/T_H} $ (straight line) with $T_H =2.1 /r_0 $. The
  smoothed cumulative number with $r_0 \Delta M = 0.5 $ is also shown. 
}
\label{fig:logNcum}
\end{figure}

The known glueball spectrum as obtained from the lattice ends at a
maximum mass value $M_{\rm max}$. Although this is somewhat tangential
to the subject of the present paper it is interesting to
analyze the effects due to a possible contribution of higher mass
states. The density of states is defined as 
\begin{eqnarray}
\rho (M) = \sum_i g_i \delta (M-M_i) = \frac{dN(M)}{dM} , 
\end{eqnarray}
where $N(M)$ is the cumulative number of glueball states
\begin{eqnarray}
N (M) = \sum_i g_i \Theta (M-M_i) .
\end{eqnarray}
Here we conventionally take $\Theta (0)=1/2$. A logarithmic plot of
$N(M)$ for the glueball spectrum is presented in
Fig.~\ref{fig:logNcum} where a rough straight line can be envisaged
with the exception of low lying states and the highest states,
presumably due to boundary effects.\footnote{For the skeptical reader
we note that recent updates of the particle spectrum provide a similar
pattern for both mesons and baryons separately~\cite{
Broniowski:2000bj,Broniowski:2004yh}.}  We fit the function
\begin{eqnarray}
N(M) = A e^{M/T_H}  
\end{eqnarray}
by minimizing 
\begin{eqnarray}
\chi^2 = \sum_i \left(\log N(M_i) - \log( A e^{M_i/T_H}) \right)^2
.
\end{eqnarray}
This gives $r_0 T_H =2.1$ or equivalently $T_H= 2.8 \, T_c$. A weak
point of the present determination is the inability to provide a
reliable error estimate for the former fit, as there are fluctuations
in the spectrum that our smooth fitting function can never account
for. To consider this possibility we also represent the {\it
smoothed} cumulative number
\begin{eqnarray}
\langle N_{\rm lat}(M) \rangle  =  \sum_i g_i \left(
\frac{1}\pi \tan^{-1} \!\left[ 
\frac{M-M_i}{\Delta M}
\right] 
+\frac{1}{2}
\right)
\end{eqnarray}
which mimics a Breit-Wigner finite width in the density of
states. Fluctuations in the spectrum are washed out already with a
common smoothing $r_0 \Delta M  =0.5 $, a factor only 5 times larger than
the uncertainty in the lowest glueball mass. As we can see, the smoothing
tends to confirm the fit, suggesting the robustness of the analysis.

\section{Perturbative expansion of the free energy}
\label{app:3}

We compute in this appendix the perturbative expansion of the free energy of a hot gluon plasma using a renormalization group (RG) improvement of the series.  The pressure of gluodynamics has been calculated within the weak coupling expansion
through ${\cal O} (g^6 \log g ) $ in
Ref.~\cite{Shuryak:1977ut,Chin:1978gj,Kapusta:1979fh,%
  Toimela:1982hv,Arnold:1994eb,Zhai:1995ac,Braaten:1995cm,Kajantie:2002wa},
up to an unknown constant. The result can be written in the form
\begin{eqnarray} 
H_{\rm pert} 
&=& 
1 + a_2 \alphac
 + a_3 \alphac^{3/2} 
\nonumber \\ 
&&
+ \alphac^2 \left[ a_4 +
b_4 \log \left(\frac{\mu}{2\pi T}\right) 
+ c_4 \log\alphac \right] 
\nonumber \\ 
&&
+ \alphac^{5/2}\left[ a_5 + b_5 \log
\left(\frac{\mu}{2\pi T}\right) \right] 
\nonumber \\ 
&& 
+\alphac^3 \Big[ a_6 + b_6 \log
\left(\frac{\mu}{2\pi T}\right) + c_6 \log\alphac
\nonumber \\ 
&&
+ d_6 \log\left(\frac{\mu}{2\pi T}\right) \log\alphac
+ e_6 \log^2\left(\frac{\mu}{2\pi T}\right) \Big]
\nonumber \\ 
&& +{\cal O}(\alphac^{7/2}),
\label{eq:Hpert}
\end{eqnarray} 
where we have defined $H_{\rm pert} = P_{\rm pert}/P_{\rm ideal}$,
being $P_{\rm ideal} = ((N_c^2-1)\pi^2/45) T^4$ the pressure of an
ideal gas of massless gluons. In this formula $\alphac= \alphac(\mu)=
g^2(\mu)/4\pi$ is the running coupling constant. All coefficients
appearing in (\ref{eq:Hpert}) are known except $a_6$, which cannot be
computed within the perturbative scheme owing to infrared divergencies
\cite{Linde:1980ts}.

The trace anomaly follows straightforwardly from the expression of
$((N_c^2-1)\pi^2/45)H_{\rm pert}$ by applying $T \partial/\partial T$
(the derivative affects only the explicit $T$ dependence).

We can see in Eq.~(\ref{eq:Hpert}) that $P_{\rm pert}(T)$ is computed
in the weak coupling limit as an expansion in powers of $\alphac(\mu)$
with coefficients which depend on $\log(\mu/2\pi T)$. However the
dependence in $\mu$ disappears when all orders are added
\begin{eqnarray}
\mu \frac{d }{d\mu} H_{\rm pert} = 0 .
\end{eqnarray}
This is because not all coefficients $a_i$, $b_i$, $c_i$, are independent.
Indeed
\begin{eqnarray}
\mu \frac{\partial \alphac(\mu)}{ \partial \mu } 
= \betaa (\alphac) \label{eq:betaf}
\end{eqnarray} 
with
\begin{eqnarray}
\betaa(\alphac) = - \alphac^2 \left( \betaa_0 + \betaa_1 \alphac 
+ \betaa_2 \alphac^2  + \cdots  \right) 
\label{eq:beta_expansion}
\end{eqnarray} 
and $\betaa_0 = \frac{22}{4\pi} $, $\betaa_1 = \frac{204}{(4 \pi)^2} $
and $\betaa_2 = \frac{2857}{(4 \pi)^3}$ for $N_c=3$.
When this is applied to Eq.~(\ref{eq:Hpert}) it implies
\begin{eqnarray} 
b_4 &=& \betaa_0 a_2 
,
 \nonumber \\ 
b_5 &=& \frac{3}{2} \betaa_0 a_3
,
 \nonumber \\ 
b_6 &=& \betaa_0 (2 a_4 + c_4) + \betaa_1 a_2
,
\nonumber \\
d_6 &=& 2 \betaa_0 c_4 ,
\nonumber \\
e_6 &=& \betaa_0^2 a_2 
\,.
\label{eq:bs}
\end{eqnarray} 
Of course, RG invariance is spoiled if only a
finite number of terms is retained. For the truncated series to be
useful, $\alphac(\mu)$ must be small, i.e., $\mu \gg \Lambda_{\rm
QCD}$, with fixed $2\pi T/\mu$, which in turn requires $T \gg
\Lambda_{\rm QCD}$.

We can go further and try to reorganize the series of $H_{\rm pert}$
in such a way that it appears explicitly RG invariant. To this end we
need to identify some invariant quantity and make the expansion around
it. The solution of Eq.~(\ref{eq:betaf}) can be written as
\begin{eqnarray}
-\int^{\alphac(\mu_0)}_{\alphac (\mu)} \frac{d
\alphac}{\betaa(\alphac)} =\log \left( \frac{\mu}{\mu_0} \right) \,.
\end{eqnarray} 
Defining, as usual, $\Lambda_{\rm QCD}$ by the condition
 $\alphac (\Lambda_{\rm QCD}) = \infty$, implies
\begin{eqnarray}
-\int^\infty_{\alphac(\mu)} 
\frac{d \alphac}{\betaa(\alphac)} 
&=&
\log \left( \frac{\mu}{\Lambda_{\rm QCD}} \right) 
\label{eq:6.7}
 \\
&=&
\log \left( \frac{2\pi T}{\Lambda_{\rm QCD}} \right) 
+ \log \left( \frac{\mu}{2\pi T} \right) 
.
\nonumber
\end{eqnarray} 
This allows to express $\alphac(\mu)$, appearing in the left-hand
side, in terms of $\log(\Lambda_{\rm QCD}/2\pi T)$ and $\log(\mu/2\pi
T)$.  For convenience, we introduce the quantity
\begin{equation}
\R(T) = \frac{1}{\betaa_0} \frac{1}{\log\left(\frac{ 2 \pi
T}{\Lambda_{\rm QCD}}\right)} \,,
\label{eq:appRGinv}
\end{equation}
which is manifestly RG invariant and behaves as $\alphac(\mu)$ when
$\mu=2\pi T\to \infty$. From (\ref{eq:6.7}) and
(\ref{eq:beta_expansion}) then
\begin{eqnarray}
\alphac(\mu) 
&=& 
\R
+ \R^2 \left[
 \frac{\betaa_1}{\betaa_0} \log\R 
- \betaa_0 \log\left( \frac{\mu}{2 \pi T}\right)
\right]
\nonumber \\
&&
+ \R^3 \bigg[ 
 \frac{\betaa_2}{\betaa_0} -\frac{\betaa_1^2}{\betaa_0^2} 
+ \frac{\betaa_1^2}{\betaa_0^2} \log\R 
+ \frac{\betaa_1^2}{\betaa_0^2} \log^2\R 
\nonumber \\
&&
-  \betaa_1 \log\left(\frac{\mu}{2 \pi T}\right) 
+ \betaa_0^2 \log^2\left( \frac{\mu}{2 \pi T}\right)
\nonumber \\
&&
- 2 \betaa_1 \log\R \log\left(\frac{\mu}{2 \pi T}\right) 
\bigg] 
\nonumber \\
&&
+ {\cal O}(\R^4) 
\,.
\label{eq:Rexpansion}
\end{eqnarray} 
By introducing this perturbative series of $\alphac(\mu)$, in
Eq.~(\ref{eq:Hpert}) one obtains
\begin{eqnarray} 
H_{\rm pert} &=& 1 + A_2 \R + A_3 \R^{3/2}
\nonumber \\ 
&& + 
\R^2 \left[ A_4 + B_4 \log \R \right] 
\nonumber \\
&& + 
\R^{5/2} \left[ A_5 + B_5 \log \R \right] 
\nonumber \\ 
&& + 
\R^3
\left[ A_6 + B_6 \log \R + C_6 \log^2 \R \right]
\nonumber \\ 
&& +
{\cal O}(\R^{7/2})
.
\label{eq:pRG}
\end{eqnarray}
All coefficients here are known except $A_6$. In fact\footnote{The values for the coefficients are 
$A_2 = -15/(4\pi)$, 
$   A_3 = 30/\pi^{3/2}$,
$   A_4 = 135( 3.51 - \log\pi )/(2\pi^2)$,
$   B_4 = 5175/(88\pi^2)$,
$   A_5 = -3.23*495/(2\pi^{5/2})$,
$   B_5 = 2295/(22\pi^{5/2})$,
$   B_6 =  2295/(1936 \pi^3)(115 + 264(3.51 - \log\pi )) - 1134.8/\pi^3$,
$   C_6 = 566865/(1936\pi^3)$.
}
\begin{eqnarray} 
A_2 &=& a_2 ,
 \nonumber \\
A_3 &=& a_3 ,
 \nonumber \\
A_4 &=& a_4 ,
\nonumber \\
B_4 &=& \frac{\betaa_1}{\betaa_0} a_2 + c_4 ,
\nonumber \\
A_5 &=& a_5 ,
\nonumber \\
B_5 &=& \frac{3}{2}\frac{\betaa_1}{\betaa_0} a_3 ,
 \nonumber \\
A_6 &=& \left( \frac{\betaa_2}{\betaa_0} - \frac{\betaa_1^2}{\betaa_0^2}\right) a_2 + a_6 ,
\nonumber \\
B_6 &=& \frac{\betaa_1^2}{\betaa_0^2} a_2 + 2 \frac{\betaa_1}{\betaa_0} a_4 + \frac{\betaa_1}{\betaa_0} c_4 + c_6 ,
\nonumber \\
C_6 &=& \frac{\betaa_1^2}{\betaa_0^2} a_2 + 2 \frac{\betaa_1}{\betaa_0} c_4   
. \label{eq:RGcoef}
\end{eqnarray} 

$H_{\rm pert}$ is now written in a manifestly RG invariant way as a
function of $T/\Lambda_{\rm QCD}$. In principle, the parameter
$\Lambda_{\rm QCD}$ itself can be extracted from the pressure at
sufficiently high temperature:
\begin{eqnarray}
\log\Lambda_{\rm QCD}
&=&
\frac{1}{\betaa_0 A_2}\lim_{T\to\infty}
\Big(
\frac{H_{\rm pert}-1 }{t^2}
-\frac{A_2}{t}
\nonumber \\ &&
-\frac{ A_3}{ t^{1/2}}
- B_4 \log t 
-A_4
\Big)
,
\end{eqnarray}
where $t=1/(\betaa_0\log(2\pi T))$ (and $\Lambda_{\rm QCD}$ results in
the same units used for $T$ in $t$). In practice, the pressure is not
known to the required precision to such high temperatures and the
parameter $\Lambda_{\rm QCD}$ is obtained from other determinations.

The trace anomaly follows immediately from the RG expression noting that
$d\R/d\log T = -\betaa_0 \R^2$,
\begin{equation}
\Delta_{\rm pert} = 
-\frac{(N_c^2-1)\pi^2}{45}\betaa_0\R^2\frac{d H_{\rm pert}}{d\R}
\,, \label{eq:appDeltaPT}
\end{equation}
known modulo ${\cal O}(\R^{9/2})$.



\end{document}